\documentclass[aps,twocolumn,showpacs]{revtex4}

  \usepackage{graphicx,natbib}
  \usepackage{bm,amsmath,amsfonts,color}

    \newcommand{\bz} {{\bf z}^{\pm}}
    \newcommand{\zp} {z_{i}^{\pm}}
    
    \newcommand{\dz} {\delta z_{i}^{\pm}}

        \newcommand{\uv}[1] {\hat{ {\bf #1}}}   %.. Unit Vector

        \renewcommand{\d} {\mathrm{d}}          %.. for use in
\begin{document}

\title{The third-order law for increments in magnetohydrodynamic
    turbulence with constant shear}

\author{M.  Wan$^1$,
    S.  Servidio$^1$,
    S.  Oughton$^2$,
    and
    W.  H.  Matthaeus$^1$}

\affiliation{$^{1} $Bartol Research Institute and Department of Physics
    and Astronomy,
    University of Delaware,
    Newark, Delaware 19716
    \\$^{2} $Department of Mathematics,
        University of Waikato,
        Hamilton, New Zealand}

\begin{abstract}
We extend the theory
for third-order structure functions
in homogeneous incompressible magnetohydrodynamic (MHD) turbulence to
the case in which a constant velocity shear
is present.
A generalization is found of the
usual relation
    [Politano and Pouquet, Phys.  Rev.  E, \textbf{57} 21 (1998)]
between third-order structure functions and the
dissipation rate in steady inertial range
turbulence, in which the shear plays a crucial role.
In particular, the presence of shear leads to a third-order law which
is \emph{not} simply proportional to the relative separation. 
Possible implications for laboratory and space plasmas are discussed.
\end{abstract}

\pacs{XXX}

\maketitle

    \section{Introduction}
A well known result in hydrodynamic turbulence theory is the
Kolmogorov--Yaglom (``4/5'')
law that relates the third-order structure
function to the energy dissipation rate
        \cite{Kol41c,MoninYaglom,Frisch}.
Often regarded as a
rigorous result of the fluid equations, this law
requires assumptions of isotropy, homogeneity, and time stationarity of the
statistics of velocity increments
    $ \delta {\bf u} =
        {\bf u}({\bf x} + {\bf r}) - {\bf u}({\bf x}) $
(velocity ${\bf u}$,
spatial positions ${\bf x} + {\bf r}$ and ${\bf x}$).
In addition, and crucially, 
it also requires adoption of the von K\'arm\'an hypothesis \cite{KarmanHowarth38}
that the rate of
energy dissipation $\epsilon$ approaches a constant nonzero value
as Reynolds number tends to infinity.  Without the need for
assuming isotropy, one finds
\begin{equation}
  \frac {\partial}{\partial r_i}
    \langle
    \delta u_i |\delta {\bf u}|^2
    \rangle
    =
    -4 \epsilon,                    \label{eq:3rdorderDIV}
\end{equation}
where $\langle \cdots \rangle$ indicates an ensemble average
and a sum on repeated indices is implied.
If isotropy is further assumed then,
 \begin{equation}
  \langle \delta u_L |\delta {\bf u}|^2 \rangle
    = -\frac{4}{d}
           \epsilon |{\bf r}| ,                 \label{eq:3rdOrderISO}
\end{equation}
where $d$ is the number of spatial dimensions and
        $ \delta u_L = \hat {\bf r} \cdot \delta {\bf u}$
is the increment component measured in the direction of the 
unit vector $\uv{r}$ 
parallel to the relative separation ${\bf r}$.
Extension of the third-order law to the
case of incompressible MHD was reported by
        \citet{PolitanoPouquet98},
who remained close to the approximations
made in the hydrodynamic case.  Without assuming isotropy, they
found
\begin{equation}
   \frac{\partial}{\partial r_k}
    \langle
            \delta z_k^{\mp}   | \delta {\bf z}^\pm |^2
    \rangle
      =
        - 4 \epsilon^{\pm},             \label{eq:3rd_diff}
\end{equation}
which, after adoption of isotropy, reduces to,
\begin{equation}
   \langle
    \delta{z_L^{\mp}}|\delta {\bf z^\pm}|^2
   \rangle
    =
    - \frac{4}{d} \epsilon^{\pm} r,
                        \label{eq:IsoMHDLaw}
\end{equation}
where
    $ \delta {\bf z}^\pm
         =
     {\bf z}^\pm({\bf x+\bf r})- {\bf z}^\pm(\bf x)$
are the increments of the Els\"asser
variables and $\delta z_L^\pm = \uv{r} \cdot \delta {\bf z}^\pm$.
The constants $\epsilon^{\pm}$ are the mean energy dissipation
rates of the corresponding variables
    $ {\bf z}^\pm= {\bf u} \pm {\bf b}$,
where $\bf b$ is the magnetic field
fluctuation in Alfv\'en speed units.

Here we extend the third-order law in MHD
turbulence to cases in which the isotropy assumption is relaxed.
This is accomplished by introducing homogeneous shear
in the velocity field,
a simplified and well-studied approach in hydrodynamics
    \cite{Rogallo81,LeeEA90,KidaTanaka94,Pumir96}.
In particular, it supports
  {departures from strict isotropy and}
introduction of shear without
consideration of rigid boundaries.
MHD third-order laws have been
applied to systems
that may also admit departures
from strict uniformity, due to coherent
large-scale gradients;
        e.g.,
plasma confinement devices \cite{SerianniEA07,LepretiEA09}
and the
solar wind
    \cite{TuMarsch95,Sorriso-ValvoEA07,MacBrideEA08,MarinoEA08}.
For systems like these,
the homogeneous shear approximation may be a reasonable
step towards including such large-scale effects in the
relevant MHD turbulence scaling laws.  
To this end, our derivation of the MHD third-order law will include
the effect of homogeneous shear, 
leading to a necessarily anisotropic form for the law.

More specifically, 
we find that a uniform shear introduces new terms in the
third-order law, so that one can no longer conclude that a
particular third-order structure function, or even a particular
integral of a third-order structure function, is proportional to
the dissipation rate times the relative separation length $r$.
 This is in
marked contrast to the situation for the fully isotropic hydrodynamic
and MHD cases, given here as
    Eqs.~(\ref{eq:3rdOrderISO}) and (\ref{eq:IsoMHDLaw}).
It is, however,
entirely consistent with the work of
    \citet{Lindborg96}
and
    \citet{CasciolaEA03},
who derived modifications to the form of the
third-order law for hydrodynamics with shear.

The principle theoretical result given below is that a
uniform shear indeed is responsible for changing the form of the
third-order law, whereas a mean magnetic field does not produce
such structural changes.  Implications for solar wind, 
laboratory, and 
astrophysical measurements of turbulence are suggested,
and in particular the primacy of the third-order law in
unambiguously defining an inertial range is challenged.

    \section{Energy decay with large-scale fields}

% Following the same procedure used to obtain the 4/5-law for
% MHD
%     \cite{PolitanoPouquet98,Sorriso-ValvoEA07},
% and combining with the extension of Eq.~(\ref{eq:3rdOrderISO})
% performed by
%     \citet{CasciolaEA03},
% we will obtain the MHD third-order law in
% the presence of a constant shear of velocity.

The third-order law is often derived from the steady-state version of an
equation related to energy decay.
To obtain the version of the law appropriate for MHD with uniform
velocity shear, 
we follow the same procedure used previously for MHD 
    \cite{PolitanoPouquet98,Sorriso-ValvoEA07},
combined with the method
of \citet{CasciolaEA03} for
extending Eq.~(\ref{eq:3rdOrderISO})
to include shear. 
A uniform magnetic field is also retained,
although only the simplest of its consequences
will enter the discussion.

First, let us employ a Reynolds decomposition of the
velocity field
    ${\bf v}={\bf U}+{\bf u}$
into a mean velocity
    ${\bf U} ({\bf x}) $
and a fluctuating component
    $ {\bf u} ({\bf x},t) $,
where
    $ \langle {\bf v} \rangle = {\bf U}$ and
    $ \langle {\bf u} \rangle = \boldmath{0}$.
Similarly we write the total magnetic field, conveniently
expressed in Alfv\'en speed units, as
    ${\bf B} = {\bf b} + {\bf B}_0$.
We assume ${\bf B}_0$ is constant and uniform, but that
    ${\bf U} ({\bf x})$
varies in space.  However this variation will be taken as
non-random and slowly-varying, so that the turbulence properties
can be treated as locally homogeneous.  
%% In the next section we refine this further.

Now we write the incompressible MHD equations at two positions,
        $\bf x$ and ${\bf x}^{\prime}={\bf x}+{\bf r}$:
\begin{equation}
      \partial_t{\zp}
    =
    - (z_{k}^{\mp}+U_k \mp B_{0k})\partial_k(U_i+\zp)
    - \partial_i{P}
    + \nu\partial_k\partial_k\zp,
                            \label{eq:zx}
\end{equation}
\begin{equation}
    \partial_t{z_{i}^{\pm\prime}}
  =
    -
      ( z_{k}^{\mp\prime} + U_{k}^{\prime} \mp B_{0k} )
       \partial_k^{\prime} (U_{i}^{\prime} + z_{i}^{\pm\prime})
    -
      \partial_i^{\prime}{P^{\prime}}
    +
      \nu\partial_k^{\prime}\partial_k^{\prime}z_{i}^{\pm\prime}.
                            \label{eq:zxp}
\end{equation}
Here the prime denotes quantities at position ${\bf x}^{\prime}$, $P$
is the pressure, and $\nu$ is the kinematic
viscosity, taken equal to the resistivity hereafter.
Subtracting
    Eq.~(\ref{eq:zx}) from Eq.~(\ref{eq:zxp})
yields the following equation for the
Els\"asser  increments
    $\delta\bz=\bz({\bf x}^{\prime})-\bz(\bf x)$:
\begin{eqnarray}
  \partial_t{\delta \zp}
    & = &
    -  ( \delta{U_k} + \delta{z_k^{\mp}})
                \partial_k^{\prime}\delta\zp
   \nonumber \\
        & &
        -  ( z_k^{\mp} + U_k \mp B_{0k})
            (\partial_k^{\prime} + \partial_k) \delta\zp
   \nonumber \\
        & &
        -  (\delta{z_k^{\mp}} + \delta U_k)
            \partial_k{U_i}\nonumber
   \\
    & &
    - (z_k^{\mp \prime} + U_k^{\prime} \mp B_{0k})
        \delta (\partial_k U_i)
   \nonumber \\
    & &
    - (\partial_i^{\prime}+\partial_i)\delta P
    + \nu (\partial_k^{\prime}\partial_k^{\prime} + \partial_k\partial_k)
        \delta\zp
     ,
  \label{eq:deltaZ}
\end{eqnarray}
where we use the property that the
primed and unprimed coordinates are independent, so that
    $ \partial_k{z_i^{\pm\prime}}= 0 $
and     $ \partial_k^{\prime}\zp = 0 $.

As noted above, we seek an equation related to energy decay.
Multiplying the previous equation
by $2\delta\zp$ and averaging yields
\begin{eqnarray}
  \partial_t\langle\lvert\dz\rvert^2\rangle 
        & = &
      -\frac{\partial}{\partial{r_k}}
          \langle( \delta U_k+\delta z_k^{\mp}) \lvert\dz\rvert^2
          \rangle
   \nonumber \\
        & &
      +   \langle 
                \lvert \dz \rvert^2 
                 (\partial_k {U_k} + \partial_k^{\prime}{U_k^{\prime}})
          \rangle
   \nonumber \\
        & &
          -2 \langle
                \partial_k{U_i}\dz(\delta{z_k^{\mp}}+\delta{U_k})
             \rangle
   \nonumber \\
        & &
          -2 \langle
                ( z_k^{\mp \prime} +U_k^{\prime} \mp B_{0k})
                \delta (\partial_k U_i) \dz
             \rangle
   \nonumber \\ 
        & &
          + 2 \nu \frac{\partial^2}{\partial{r_k}^2}
                \langle 
                        \lvert \dz \rvert^2 
                \rangle
          - 4 \nu 
                \langle 
                        \lvert \partial_k \zp \rvert^2 
                \rangle.
  \label{eq:deltaZ2_ave}
\end{eqnarray}
In arriving at this expression we make use of
    $ \partial_k\langle\bullet\rangle
      =
          -\frac{\partial}{\partial{r_k}}\langle\bullet\rangle $
and
        $ \partial_k^{\prime}\langle\bullet\rangle
      =
       \frac{\partial}{\partial{r_k}}\langle\bullet\rangle $.
These latter relations follow from spatial homogeneity
(i.e., translation invariance of the statistical properties),
which can be considered for some systems to be 
an exact property (see following section)
or an approximation, e.g., in the case of a weakly inhomogeneous 
system. The main results here will be for strict homogeneity. 

The last term of Eq.~(\ref{eq:deltaZ2_ave}) 
can be identified with the
dissipation rates
\begin{equation}
    \epsilon^{\pm}= \nu\langle \lvert \partial_k \zp \rvert^2 \rangle,
\label{eq:epsilon}
\end{equation}
which for steady state are also the mean energy transfer rates.
Following the usual arguments \cite{KarmanHowarth38}, in the
limit of vanishing viscosity $\nu \to 0$, 
it is assumed---not proven---that 
        the $\epsilon^{\pm}$ remain nonzero, 
and in effect are
externally prescribed by the rate of supply of turbulence energy 
(and cross helicity).
Although this nontrivial assertion is physically plausible
    \cite{EyinkSreenivasan06}, 
it nonetheless prevents the
% This is a nontrivial, although physically plausible
% assertion
%     \cite{EyinkSreenivasan06},
% which however prevents the
subsequent developments, including the classical 4/5-law, from
being considered an exact consequence of the fluid equations
themselves.   
Furthermore, the penultimate  
term in
        Eq.~(\ref{eq:deltaZ2_ave}), 
also involving the viscosity, is
assumed to vanish at high Reynolds number when we are examining
the inertial range of separations. For the above-stated set of
approximations, the increments $r$ are restricted to lie in the
inertial range, that is separations smaller than the correlation
length (energy-containing scale) and bigger than the dissipation
scale (scale at which fluctuations are critically damped).
For variations of the set of assumptions that lead to
a third-order law,
    see e.g., \citet{Hill97}.

    \section{MHD with homogeneous shear}

The above relations need not be strictly
homogeneous, as variations in
%% the large-scale velocity
    $ {\bf U} $
over the slowly-varying large scales may be present.
To rectify this and
arrive at a general law that is translation invariant, we now
specialize to the case of a homogenous shear flow, alluded to in
the introductory section.
With this choice the tensor
    $\partial U_i/\partial x_j$ is a constant matrix
independent of position.
The turbulence is then homogeneous and
all terms in
    Eq.~(\ref{eq:deltaZ2_ave})---both coefficients and averaged
terms---are only a function of the separation vector $\bf r$.

Under the hypothesis of steady-state turbulence, 
the left-hand side of Eq.~(\ref{eq:deltaZ2_ave}) vanishes. 
Integrating in $\bf r$, over a volume $\cal V$ that is
enclosed by a surface $\cal S$, the equation becomes:
\begin{eqnarray}
& &
\oint_{\cal S}
  \left[
        \hat{n}_k
    \langle (\delta{z_k^{\mp}} + \delta{U_k} )
        \lvert \dz \rvert^2
    \rangle
  \right]
     \d S_r
  +
      2\frac{\partial U_i} {\partial x_k}
     \int_{\cal V}
    \langle
            \dz \delta{z_k^{\mp}}
    \rangle
       \, \d V_r
  \nonumber \\
   & & =
    - {4 V }\epsilon^{\pm},
                        \label{eq:3rd_int_general}
\end{eqnarray}
where $V$ is the volume of the region $\cal V$ 
and $\hat{n}_k$ is a unit vector normal to $\cal S$.

If the
region of integration is
a three dimensional sphere of radius $r$, volume $V_r$
and surface $S_r$, the integration
yields
\begin{eqnarray}
& &
  \oint
     \left[
         \hat{r}_k
        \langle 
          ( \delta{z_k^{\mp}}+\delta{U_k} ) \lvert \dz \rvert^2
        \rangle
     \right]  \d S_r
      +
        2\frac{\partial U_i}{\partial x_k}
    \int
       \langle\dz\delta{z_k^{\mp}} \rangle \d V_r
   \nonumber \\
    & &
     =
       - \frac{16\pi r^3}{3} \epsilon^{\pm},
  \label{eq:3rd_int}
\end{eqnarray}
where, in the first term of the equation, $\hat{r}_k$ is the unit
vector normal to the surface of the sphere, and now in spherical
    $(r,\theta, \phi)$
coordinates
  $\d S_r = r^2 \d(\cos{\theta}) \d{\phi} \equiv r^2 \d\Omega$.
Equation~(\ref{eq:3rd_int})
may be interpreted as the integral form of
the third-order law for incompressible homogeneous MHD turbulence
with an external velocity field that is constant in time but which
can vary linearly in space.
By setting ${\bf U}=\boldmath{0}$ and assuming
isotropic turbulence,
    Eq.~(\ref{eq:3rd_int})
will recover the
standard third-order law for isotropic MHD
turbulence
    \cite{PolitanoPouquet98},
given here as
    Eq.~(\ref{eq:IsoMHDLaw}).

In standard derivations for isotropic turbulence
    \cite{PolitanoPouquet98,Sorriso-ValvoEA07,MacBrideEA08},
shear is
necessarily lacking, and it is assumed that the structure
functions are rotationally symmetric.  In that case the above
relation is simplified by carrying out the integrals explicitly.
(For a more general case, see below.)
Here we allow for anisotropy induced by either the
large-scale magnetic field, or by the imposed homogeneous shear.
Note that the large-scale magnetic field ${\bf B}_0$
does not appear explicitly in the third-order relation, even
though it is well documented that such a field induces spectral
anisotropy in  MHD turbulence \cite{ShebalinEA83}.

We now further specialize to the
large-scale homogeneous
shear flow
    $ {\bf U} =  U_x(y) \uv{x} = \alpha y \uv{x}$
in a cartesian
$(x,y,z)$ system, with $\alpha = const.$ The integral form of the
third-order relation becomes
\begin{eqnarray}
  & &
    r^2 \oint
    \left [ \uv{r} \cdot
          \langle
         ( \delta{{\bf z}^{\mp}}\lvert \delta {\bf z}^\pm\rvert^2
          \rangle
        \right]  \d\Omega
  \nonumber \\
    & + &
    \alpha r^3
    \oint
        (\uv{r} \cdot \uv{x}) (\uv{r} \cdot \uv{y})
            \langle \lvert \delta {\bf z}^\pm \rvert^2 \rangle
            \, \d\Omega
  \nonumber \\
    &+ &
    2 \alpha
    \int
       \langle \delta z_x^\pm \delta{z_y^{\mp}} \rangle
        \, \d V_r
    =
          - \frac{16\pi r^3}{3} \epsilon^{\pm} .
                                \label{eq:3rd_int_2}
\end{eqnarray}
Denoting an angular average over a shell of
radius $r$ as $\langle \dots \rangle_{\Omega}$ and a volume average
over a sphere of radius $r$ as   $\langle \dots \rangle_V$ we may
rearrange the above equation as
\begin{eqnarray}
  \langle \langle
    \delta{{z}_L^{\mp}}\lvert \delta {\bf z}^\pm\rvert^2
  \rangle \rangle_{\Omega}
 & = &
    - \alpha r
        \langle \langle
            (\uv{r} \cdot \uv{x})
            (\uv{r} \cdot \uv{y})
            \lvert \delta {\bf z}^\pm \rvert^2
        \rangle \rangle_{\Omega}
  \nonumber \\
& &
   -  \frac {2}{3} \alpha  r
    \langle \langle
        \delta z_x^\pm \delta{z_y^{\mp}}
    \rangle \rangle_V
   -  \frac{4}{3} r \epsilon^{\pm},
                    \label{eq:3rd_averages}
\end{eqnarray}
where, again, $\delta z_L^\pm = \uv{r} \cdot \delta {\bf z}^\pm$.  This
form, based on a spherical region of radius $r$, 
indicates
that all three terms on the right hand side of the
equation have an explicit proportionality to $r$; moreover,
the first and second of these also admit an \emph{implicit} dependence on $r$.
The quantity on
the left side of
        Eq.~(\ref{eq:3rd_averages})
is the MHD analog of
the usual third-order
structure function that appears in the Yaglom and Kolmogorov laws
        \cite{Kol41c,MoninYaglom},
and we see that
in the presence of homogeneous shear
it is \emph{not}
simply proportional to the dissipation
    $\epsilon^\pm$.

At this point
we remark on an alternative form that the third-order law
can assume that may be revealing in anisotropic cases.  Recall that
    Eq.~(\ref{eq:3rd_int_general})
is valid for an \emph{arbitrary} volume
$\cal V$ and its associated bounding surface $\cal S$.  
The
advantage of employing a spherical volume $\cal V$ is that
when the flux is isotropic, the integrand in the surface integral will
be independent of the direction of $\bf r$, 
making the integration trivial.
Unfortunately, this property is lost when the turbulence is
anisotropic \cite{ShebalinEA83,TeacaEA09}.
However, provided that the (energy-like) vector flux
  $ {\bf F}^+   = \langle
             ( \delta{{\bf z}^{-}} + \delta{\bf U} )
         \lvert \delta {\bf z}^+ \rvert^2
        \rangle $
is smoothly varying in $\bf r$,
it is in
principle possible to find a set of nested surfaces  $\cal S(V)$
  [labeled by their enclosing volume $V$ and with 
   unit normal vectors
      $\uv{n}_{\cal S} $],
such that the normal component of the vector flux
    $ \bf F^+$
is uniform across  $\cal S(V)$.
Then
    $ \oint_{\cal S} \d S \, \uv{n}_{\cal S} \cdot {\bf F}^+
      = F_n^+(V) S $,
where the constant normal flux $F_n$ is labeled by the volume $V$
bounded by the surface, and $S$ is the value of the surface area.  
The partner quantity 
   ${\bf F}^-$ is defined analogously.
Provided these nested surfaces can be found,
the homogeneous shear case, Eq.~(\ref{eq:3rd_int_general}), can then be
reduced to
\begin{eqnarray}
 F_n^\pm(V^\pm)
 & =&
    \langle
        \uv{n}^\pm_{{\cal S}^\pm} \cdot
        (\delta {\bf z}^\mp + \delta {\bf U} )
        \lvert \delta {\bf z}^\pm \rvert^2
    \rangle
\nonumber \\
 & = &
   -  \frac{2 \alpha V^\pm} {S^\pm}
    \langle \langle
        \delta z_x^\pm \delta{z_y^{\mp}}
    \rangle  \rangle_{\cal V}
   -    \frac{4 V^\pm}{S^\pm}  \epsilon^{\pm},
                            \label{eq:3rdOrder-SV}
\end{eqnarray}
where $V^\pm$ and $S^\pm$ are the volumes and associated
surface areas that admit constant normal fluxes $F^\pm_n (V^\pm)$.  
Note that in general the constant flux 
surfaces $S^+$ and $S^-$ are expected to be different,
e.g., due to cross helicity effects.

When
homogeneous shear is absent the result in 
        Eq.~(\ref{eq:3rdOrder-SV}) 
reduces to the formal anisotropic
third-order law
\begin{equation}
         F_n^\pm(V^\pm) = -\frac{4 V^\pm}{S^\pm}
            \epsilon^{\pm}.
                        \label{eq:new}
\end{equation}
The latter can have application 
in the cases
in which anisotropy is present due to 
a mean magnetic field ${\bf B}_0 \ne \boldmath{0}$.

     \section{Summary and discussion}\label{sec:summary}

We examined the mixed third-order Els\"asser structure functions
for MHD turbulence, incorporating a constant 
%% and uniform 
sheared velocity
(homogeneous shear)
field in addition to homogeneous fluctuations, under a set of
assumptions that parallels those used in standard turbulence
theory to derive the Kolmogorov 4/5-law.  In analogy to the
findings of
    \citet{CasciolaEA03}
and     \citet{Lindborg96}
for hydrodynamics, we find that a law can be
obtained for stationary homogeneous turbulence that relates third-order
structure functions and dissipation, but which also involves
additional terms.  
For MHD with a constant imposed shear, there are
shear-related terms that appear in this modified third-order law,
as in the hydrodynamic case.  On the other hand, 
a uniform magnetic field does
not appear explicitly in this relation.

On the basis of a very simple estimate we expect the new terms in
the third-order equation to be of significance when the large-scale
velocity increments are of the same order or larger than
the fluctuation increments at the same separation $r$, that is, when
$\delta U \sim \delta z$.  In some applications this condition may
be realized, and consequently
the classical third-order law is modified by these new terms.  
We suspect that
for solar wind turbulence, as well
as for laboratory devices, 
the present 
generalized form of the third-order law
will be relevant.
% that is applied to observational studies to include effects such as
% what we have suggested here. 
In particular, 
the 
modified 
MHD third-order law no longer admits
an interpretation purely 
in terms of energy transfer and dissipation,
and therefore differs from  
the isotropic case 
without shear.

Further extensions of the third-order law can also be undertaken.  For
example, by including a large-scale but non-uniform magnetic field.
This will induce additional terms in the generalization of
    Eqs.~(\ref{eq:3rd_int})--(\ref{eq:3rd_averages}).

As a final remark,
we note that the modifications of the third-order law for energy decay
that we describe here can be
anticipated in the structure of
scale-separated 
transport equations derived for MHD in a weakly inhomogeneous medium
\cite{TuMarsch90b,ZhouMatt90a}.
These two-scale
transport equations provide
a formalism for evolution of second-order
correlation functions,
and include
nonlinear decay, analogous to our third-order structure functions,
along with advection and shear terms.
On this basis, 
one could have already concluded 
that the third-order law requires modification in the
presence of large-scale shear.
The present study concentrated only on the special case of homogeneous shear,
and generalizations of the third-order law
have been found.

We expect that future studies based on numerical simulations
may provide explicit verification and examples of the relationships
we propose here. Taking into account effects like shear, observational studies may prove useful in a variety of
systems with large-scale shear flows, such as astrophysical and laboratory plasmas.

\begin{acknowledgments}
This research supported in part by the NSF Solar Terrestrial
Program under grant ATM0539995
and by NASA under the Heliophysics Theory Program
grant NASA NNX08AI47G.

\end{acknowledgments}

%% \bibliography{ag,hl,mp,qz,refs_whm}
 \newcommand{\BIBand} {and} %...... how 'and' appears in authors
  \newcommand{\boldVol}[1] {\textbf{#1}} %......................
  \newcommand{\SortNoop}[1] {} %.................................
  \newcommand{\au} {{A}{U}\ } %.....................................
  \newcommand{\AU} {{A}{U}\ } %.................................
  \newcommand{\MHD} {{M}{H}{D}\ } %..............................
  \newcommand{\mhd} {{M}{H}{D}\ } %..............................
  \newcommand{\RMHD} {{R}{M}{H}{D}\ } %...........................
  \newcommand{\rmhd} {{R}{M}{H}{D}\ } %...........................
  \newcommand{\wkb} {{W}{K}{B}\ } %..............................
  \newcommand{\alfven} {{A}lfv\'en\ } %.............................
  \newcommand{\Alfven} {{A}lfv\'en\ } %.............................
  \newcommand{\alfvenic} {{A}lfv\'enic\ } %...........................
  \newcommand{\Alfvenic} {{A}lfv\'enic\ }

\end{document}